\documentclass[usenatbib]{aa}
\usepackage{graphicx}
\usepackage[varg]{txfonts}
\usepackage[english,english]{babel}
\usepackage{amsmath}
\usepackage{amssymb,amsfonts,textcomp}
\usepackage{array}
\usepackage{supertabular}
\usepackage{hhline}
\usepackage[usenames]{color}
\usepackage{hyperref}
\hypersetup{colorlinks=true, linkcolor=black, citecolor=blue, filecolor=blue, urlcolor=blue}
\usepackage{xspace}
\bibpunct{(}{)}{;}{a}{}{,}

\def\gtrsim{\lower.5ex\hbox{$\; \buildrel > \frac \sim \;$}}

\newcommand{\hagn}{\mbox{{\sc \small Horizon-AGN}}}

\newcommand{\nh}{\mbox{{\sc \small NewHorizon}}}
\newcommand{\nhs}{\mbox{{\sc \small NewHorizon\,}}}

\newcommand{\ramses}{\mbox{{\sc \small RAMSES}}}
\newcommand{\sunset}{\mbox{{\sc \small SUNSET}}}
\newcommand{\skirt}{\mbox{{\sc \small SKIRT}}}

\newcommand{\Reff}{$R_{\mathrm{e}}$}

\newcommand{\PAo}{PA$^{\mathrm{opt}}$}
\newcommand{\PAk}{PA$^{\mathrm{kin}}$}
\newcommand{\PAgal}{PA$^{\mathrm{gal}}$}
\newcommand{\PAX}{PA$^{\mathrm{X}}$}
\newcommand{\PAY}{PA$^{\mathrm{Y}}$}

\newcommand{\dPAgalopt}{$\Delta \mathrm{PA^{\mathrm{gal-opt}}}$}
\newcommand{\dPAgalkin}{$\Delta \mathrm{PA^{\mathrm{gal-kin}}}$}
\newcommand{\dPAjetopt}{$\Delta \mathrm{PA^{\mathrm{jet-opt}}}$}
\newcommand{\dPAjetkin}{$\Delta \mathrm{PA^{\mathrm{jet-kin}}}$}

\newcommand{\dPAoptkin}{$\Delta \mathrm{PA^{\mathrm{opt-kin}}}$}
\newcommand{\dPAXY}{$\Delta \mathrm{PA^{\mathrm{X-Y}}}$}

\newcommand{\Vsig}{$V_*/\sigma_*$}

\newcommand{\sigPA}{$\sigma_{\mathrm{PA^k}}$}

\newcommand{\degrees}{$^\circ$}

\newcommand{\galactica}{\mbox{{\sc \small Galactica}}}
\newcommand{\galacticas}{\mbox{{\sc \small Galactica\,}}}

\definecolor{grey}{rgb}{0.75,0.75,0.75}
\definecolor{Orange}{rgb}{1.0,0.5,0.15}
\definecolor{brown}{rgb}{0.7,0.25,0.0}
\definecolor{pink}{rgb}{1.0,0.5,0.5}
\definecolor{darkerred}{rgb}{0,0.5,0.5}
\definecolor{darkerblue}{rgb}{0,0,0.8}
\definecolor{lightblue}{rgb}{0.12, 0.56, 1.0}
\definecolor{Blue}{rgb}{0,0.08,0.65}
\definecolor{Red}{rgb}{0.65,0.08,0.05}
\definecolor{Green}{rgb}{0.15,0.45,0.25}

\begin{document} 

\title{Statistics of the projected angles between
    the black-hole spin and the host-galaxy rotation axes
  from NewHorizon}

\titlerunning{jet-galaxy alignment in NewHorizon}
\authorrunning{S. Peirani et al.}

\author{
S\'ebastien Peirani\inst{1,2,3}
\and Yasushi Suto\inst{4,5,3}
\and Clotilde Laigle\inst{6}
\and Yen-Ting Lin\inst{7}
\and Yohan Dubois\inst{6}
\and Sukyoung K.\,Yi\inst{8} 
} 
\institute{
ILANCE, CNRS – University of Tokyo International Research Laboratory, Kashiwa, Chiba 277-8582, Japan\\\email{sebastien.peirani@cnrs.fr}
\and Kavli IPMU (WPI), UTIAS, The University of Tokyo, Kashiwa, Chiba 277-8583, Japan
\and Department of Physics, School of Science, The University of Tokyo, 7-3-1 Hongo, Bunkyo-ku, Tokyo 113-0033, Japan
\and
Research Institute, Kochi University of Technology, Tosa Yamada, Kochi
782-8502, Japan
\and
Co-creation Organization of Regional Innovation,
Kochi University of Technology, Tosa Yamada, Kochi
782-8502, Japan
\and 
Institut d'Astrophysique de Paris, CNRS and Sorbonne Universit\'e, UMR 7095, 98 bis Boulevard Arago, F-75014 Paris, France
\and Institute of Astronomy and Astrophysics, Academia Sinica, No. 1, Section 4, Roosevelt Road, Taipei 10617, Taiwan
\and Department of Astronomy and Yonsei University Observatory, Yonsei University, Seoul 03722, Republic of Korea
}

   \date{Received ...; accepted ...}

 
  \abstract
  {Understanding the alignment between AGN jets and their host galaxies is crucial for interpreting AGN unification models, jet feedback processes, and the co-evolution of galaxies and their central black holes (BH). In this study, we use the high-resolution cosmological zoom-in simulation \nh, which self-consistently evolves BH mass and spin, to statistically examine the relationship between AGN jet orientation and host galaxy structure.
  Building upon our previous work, we extend the analysis of projected (2-d) alignment angles to facilitate more direct comparisons with recent observational studies. In our methodology, galaxy orientations are estimated using optical position angles derived from synthetic DESI-LS and Euclid images, while BH spin vectors serve as proxies for AGN jet directions. From a carefully selected sample of 100 BH–galaxy systems at low redshift, we generate a catalog of 5,000 mock optical images using a Monte Carlo approach that samples random viewing angles and redshifts.
  Our results reveal a statistically significant tendency for AGN jets to align with the orientation of their host galaxies, consistent with recent observations combining Very Long Baseline Interferometry (VLBI) and optical imaging of nearby AGNs. Furthermore, we find a slightly stronger alignment when using kinematic position angles derived from synthetic MaNGA-like stellar velocity fields.
  These findings underscore the importance of combining morphological, kinematic, and polarimetric information to disentangle the complex interplay between black hole spin evolution, accretion mode, and the galactic environment in shaping the direction of relativistic jets.  
  }


   \maketitle
%
\section{Introduction}

Most galaxies host super-massive black holes (BHs) in their
central parts. While their origins remain unanswered yet, the
co-evolution of the BHs and their galaxies are supposed to be
crucial in explaining the universality and diversity in the
present-day galaxies.
BHs grow by accreting mass from their host galaxies and generate relativistic jets that interact with both the interstellar and inter-galactic media.
Those jets
regulate the star formation activities in the host galaxy
\citep[e.g.,][]{dimatteo+05,springel+05,schawinski+06,croton+06,sijacki+07,booth-schaye09,dubois+12,choi+15,kurinchi+23}. Therefore, in addition to their strong gravitational interaction,  the BHs play significant roles in reshaping the gas, stellar, and dark
matter distribution from the central to outer parts of their host galaxies \citep[e.g.,][and references therein]{peirani+08,duffy+10,martizzi+13,dubois-seb16,peirani+17,peirani19,felipe+21}.
  
A critical and intriguing aspect of this phenomenon is the alignment
between black hole jets and their host galaxies. Indeed, the
orientation of a jet relative to a galaxy's stellar disk or
large-scale structure can provide important insights into the
processes governing BH spin, accretion disk orientation, galaxy
mergers, and the cosmic history of angular momentum transfer.
So far, the observational results are rather
diverse and do not offer a clear picture.  The jet directions in
elliptical galaxies and Seyfert hosts seem to be randomly distributed
with respect to their host galaxies’ stellar or gas rotation axes \citep{kinney+00,schmitt+02,gallimore+06}.
On the other hand, 
Several statistical analyses reported 
a correlation between the radio major axis and the optical minor axis in
radio-quiet passive ellipticals \citep{battye_and_browne09} 
or a preferential alignment between radio jet axes and the minor axis in some spiral galaxies and Seyfert 2 AGN \citep{saripalli+09}.

More recently, high-resolution and statistically robust studies
has sharpened the evidence for non-random jet–galaxy alignments.
\cite{zheng+24} examined a sample of over 3,600 radio-loud AGNs and found population-level evidence of alignment between radio jet orientations and the minor axes of their host galaxies.
These results lend strong support to models involving coherent accretion flows where gas inflow maintains a consistent direction over time.
Complementing this, \cite{fernandez+25}, using VLBI and optical imaging for a large sample of nearby AGNs, reported a statistically significant orthogonal alignment between parsec-scale radio jets and the major axes of their host galaxies. 
This highlights the strong interplay between supermassive black holes, their host galaxies, and their co-evolution over cosmic time.

On larger cosmic scales, \cite{jung+25} investigated the influence of the cosmic web on AGN orientation, finding a correlation between AGN jet directions and the filamentary structure of the large-scale environment, especially for central galaxies in nodes of the cosmic web \citep[see also][]{hutsmeneker+14}. This implies that both local galactic dynamics and cosmological inflow patterns may shape BH spin axes and jet orientations. Together, these studies highlight a multi-scale origin for jet–galaxy alignment, where internal disk-driven torques, secular evolution, and cosmic web anisotropies all play a role \citep[see, for example,][]{laigle+15, codis+18}.

The statistical alignment between AGN jets and their host galaxies suggests a possible physical connection between the jet-launching mechanism and the angular momentum axis of the central black hole. The observed diversity in alignment is underpinned by theoretical models that distinguish between different modes of BH fueling. In the chaotic accretion paradigm proposed by \cite{king&pringle06}, gas accretes onto the black hole in small, randomly oriented episodes, leading to frequent spin reorientations and thus random jet directions. This model explains the misalignment seen in elliptical galaxies and in many Seyferts with disrupted gas inflows.
Moreover,  \cite{hopkins+12} also predicted a weak correlation between the nuclear axis and the large-scale disk axis, using high-resolution simulations of gas inflows from galaxy to parsec
scales around AGN.
Alternatively, coherent accretion, in which gas flows maintain a consistent angular momentum axis over time, can align the BH spin with the galaxy disk. \cite{dotti+13} showed that in gas-rich disk galaxies, sustained inflow can rapidly align the BH spin with the inner disk, particularly in the presence of massive circumnuclear structures. This alignment may be aided by relativistic disk warping and the Bardeen–Petterson effect \citep{bardeen+75}, which causes the inner accretion disk to align with the BH spin axis due to Lense–Thirring precession.
The timescale of this process is believed to be much shorter than the lifetime of outflows or jets \citep{natarajan_pringle98}, suggesting that jet directions are generally not determined by the BH spin.

Evidence from X-ray observations demonstrating the ability of AGN to suppress cooling in galaxy clusters 
\citep[e.g.,][]{birzan+04,mcnamara+05,wise+07}
motivated the incorporation of BH feedback into galaxy-formation models. Early implementations first appeared in semi-analytic frameworks 
\citep[][]{bower+06,croton+06,lagos+08}
and shortly thereafter in pioneering hydrodynamical simulations
\citep[e.g.,][]{dimatteo+05,booth-schaye09}.
Since then, AGN feedback has become a standard component of cosmological simulation suites.

A number of widely used large-scale simulations adopt a single, usually thermal, feedback channel associated with high Eddington ratios (the so-called "quasar" mode) such as in
{\mbox{{\sc \small OWLS}}} \citep{schaye+10},
{\mbox{{\sc \small Magneticum}}} \citep{hirschmann+14}, 
{\mbox{{\sc \small EAGLE}}} \citep{eagle},
{\mbox{{\sc \small MassiveBlack-II}}} \citep{khandai+15},
{\mbox{{\sc \small Romulus}}} \citep{tremmel+17},
and {\mbox{{\sc \small Astrid}}} \citep{bird+22}. 
Other projects have moved toward more complex, two-mode prescriptions that alter the feedback mechanism at low accretion rates. 
{\mbox{{\sc \small IllustrisTNG}}}, for instance, transitions to a kinetic wind model inspired by advection-dominated, inflow-outflow solution \citep[ADIOS:][]{blandford&begelman99} type outflows
\citep{weinberger+17}
while the original {\mbox{{\sc \small Illustris}}} simulation injected thermal bubbles to mimic jet-inflated lobes
\citep{sijacki+07,vogelsberger+14}. 
Still others employ explicit kinetic jets
(Horizon-AGN: \citealt{hagn}; NewHorizon: \citealt{nhz}; SIMBA: \citealt{dave+19}),
in some cases supplemented by additional processes such as AGN-driven X-ray heating (e.g. in {\mbox{{\sc \small SIMBA}}}).

Underlying these implementations is a growing effort to model the sub-resolution physics of black-hole growth and energy release more realistically. Recent developments include treatments of accretion-disc structure, angular-momentum transport, and black-hole spin evolution
\citep[e.g.,][]{fanidakis,dubois+12,steinborn+15,fiacconi+18,griffin+19,Husko+22,koudmani,sala+24}.
An additional line of work explores the consequences of super-Eddington accretion, incorporating both its radiative and mechanical feedback channels, which has renewed relevance for understanding rapid early black-hole growth
\citep{rennehan+24,bennett+24,husko+25a,husko25b}

The relative orientations between the BH spin and galaxy
  rotation axes have been investigated using high-resolution
  cosmological simulations \citep{dubois+14a,beckmann+23,peirani+24}.
In particular, \cite{peirani+24} suggested that signatures of
  jet–galaxy alignment could be detectable through observational
  measurements of the projected (2-d) misalignment angles, but their
  analysis does not fully account for projection effects inherent in
  observational data. The present paper aims at revisiting their work
  in the 2-d analysis, to better bridge the gap between simulations
  and recent observational findings. This paper is structured as
  follows.  Section~\ref{sec:simu} provides a brief overview of the
  \nhs simulation and describes the numerical methodology employed in
  this work.  Section~\ref{sec:results} presents the main results,
  including statistical trends for both intrinsic spin–position angle
  misalignments and projected jet–galaxy misalignment angles.
  Section~\ref{sec:conclusions} summarizes our findings and offers
  concluding remarks.

\section{Methodology}
\label{sec:simu}

Throughout this paper, we analyse the results of the 
\nh\footnote{https://new.horizon-simulation.org/} simulation.
The details of the simulation have been described in many previous papers \cite[e.g,][]{nhz,peirani+24}, so we only summarize here its main features.

\subsection{The \nhs simulation}

\nh\, is a high-resolution zoom-in simulation from the \hagn\, simulation \citep{hagn}, 
focused on a spherical sub-volume with a radius of 10 comoving Mpc.
A standard $\Lambda$CDM cosmology was adopted with the total matter density
$\Omega_{\rm m}$ = 0.272, the dark energy density $\Omega_\Lambda$ = 0.728,
the baryon density $\Omega_{\rm b}$ = 0.045, the Hubble
constant $H_0$=70.4 km s$^{-1}$ Mpc$^{-1}$, 
the amplitude of the matter power spectrum $\sigma_8$ = 0.81
and the power-law index of the primordial power spectrum $n_{\rm s}$ = 0.967,
according to the WMAP-7 data \citep{komatsu+11}. 
The initial conditions have been generated with 
{\mbox{{\sc \small MPgrafic}}} \citep{mpgrafic} at the resolution of 4096$^3$ for \nh\, in contrast to 1024$^3$ for \hagn.
The dark matter mass resolution reaches 1.2$\times$10$^6$ M$_\odot$ compared to
8$\times$10$^7$ M$_\odot$ in \hagn. As far as star particles are concerned, their typical mass resolution is $\sim$10$^4$ M$_\odot$ for  \nh.

Both simulations were run with the \ramses\, code \citep{ramses} in
which the gas component is evolved using a second-order Godunov scheme
and the approximate Harten-Lax-Van Leer-Contact~\citep[HLLC,][]{toro}
Riemann solver with linear interpolation of the cell-centered
quantities at cell interfaces using a minmod total variation
diminishing scheme.  In \nh, refinement is performed according to a
quasi-Lagrangian scheme with approximately constant proper highest resolution of 34 pc.
The refinement is triggered in a quasi-Lagrangian manner,
if the number of DM particles becomes greater than 8, or the total
baryonic mass reaches 8 times the initial DM mass resolution in a
cell. Extra levels of refinement are successively added at $z=$ 9, 4,
1.5 and 0.25 (i.e., for expansion factor $a=$ 0.1, 0.2, 0.4 and 0.8 respectively).  The
simulation is currently completed down to redshift $z=0.18$.

It is well established that adaptive mesh refinement (AMR) codes, such as \ramses, do not conserve angular momentum exactly, particularly at refinement-level transitions. This limitation may affect simulations of rotationally supported systems. More generally, no numerical method (e.g., AMR, smoothed particle hydrodynamics, moving-mesh schemes)
preserves angular momentum perfectly. Discretisation effects, numerical viscosity, and sampling noise inevitably introduce small torques \citep{commercon+08,Hopkins15}.
Nevertheless, the angular-momentum content of simulated galaxies depends on much more than strict numerical conservation. The development of hydrodynamic instabilities plays a central role, as do stellar and AGN feedback, the structure of the interstellar medium, and its turbulent properties \citep[e.g.,][]{sijaki+12}.
Despite these challenges, the galaxies formed in NewHorizon exhibit morphologies that are broadly realistic, particularly in the case of spiral galaxies, providing qualitative support for the modelling when compared to observed galactic structural properties. The relatively modest impact of numerical resolution on galaxy spin is also illustrated in \cite{dubois+14_II}.

\nhs encompasses a wide range of sub-grid models such as, for instance, gas cooling, UV
background,  a model of star formation whose efficiency 
depends on the local turbulent Mach number and virial parameter~\citep{kimmetal17,trebitschetal17,trebitschetal20},
a model of type II supernovae based on the amount of linear
momentum injected at the adiabatic and snow-plow phase
\citep{kimm&cen14,kimmetal15} or
a model for BH mass growth and AGN feedback in alternating radio/quasar (jet/heating) mode \citep{dubois+12} coupled to a model of BH spin evolution \citep{dubois+14a}.
For this aspect, the BH spin is modeled on-the-fly in \nh\, and updated according to
the gas accretion and BH-BH mergers.
In the radio/jet mode, BHs 
power jets that continuously release mass, momentum and
energy. Bipolar jets are assumed as a cylinder of size $\Delta x$
in radius and semi-height, centered on the BH \citep{dubois+10}.
angle). The jets are launched with a speed of 10$^4$ km/s. 
Note that star formation and feedback in \nh\, are based on small-scale physics, combining theoretical models with very high–resolution simulations \citep[see][and references therein]{nhz}.
Black hole and AGN physics, however, are mainly calibrated using the  
local $M_{\rm BH}$–$M_\star$ in lower resolution ($\sim$kpc) simulations \citep{dubois+12}.

It should be stressed that
in our simulations, the gas accretion disk around the black hole is not spatially resolved. To address this, we assumed that the angular momentum of the accretion disk aligns with that of the gas measured at a distance of $4\Delta x\sim$136 pc (where $\Delta x$ is the highest spatial resolution) from each black hole. Although this scale is much larger than the true physical size of the accretion disk, it provides a reasonable approximation, under the expectation that the orientation of the angular momentum is largely preserved as gas flows inward. This assumption is supported by the high-resolution simulations of  \cite{maio+13}, who found that the angular momentum orientation is conserved down to scales of 1 pc, even in the presence of strong star formation feedback.
However, alternative perspectives exist in the literature. \cite{levine+10} reported that on scales of $\sim$100 pc, the direction of angular momentum can differ significantly from that on kiloparsec scales between 
z=4 and z=3. This variation arises mainly from infalling gas clumps whose interactions with the disk can substantially alter the angular momentum of nuclear gas. Similarly, \cite{hopkins+12} found only a weak correlation between the orientation of the nuclear axis and the larger-scale disk axis in high-resolution simulations of gas inflows from galaxy to parsec scales around AGN. Such sudden misalignments can result from massive clumps falling slightly off-axis or from gravitational instabilities. It is important to note that these studies focused on gas-rich galaxies at high redshift, where turbulent gas motions are likely to have a strong impact on angular momentum alignment.

Additionally, we have analyzed two other zoom simulations (nicknamed \galactica) focusing on isolated galaxies. For
them, we have used exactly the same physics and mass resolution as \nh\, but
they are located in different regions of \hagn\, (see, for instance,
\citealt{park+21}).

\subsection{Galaxy and black hole catalogs}
\label{subsec:catalog}

\begin{figure}
\begin{center}
\rotatebox{0}{\includegraphics[width=\columnwidth]{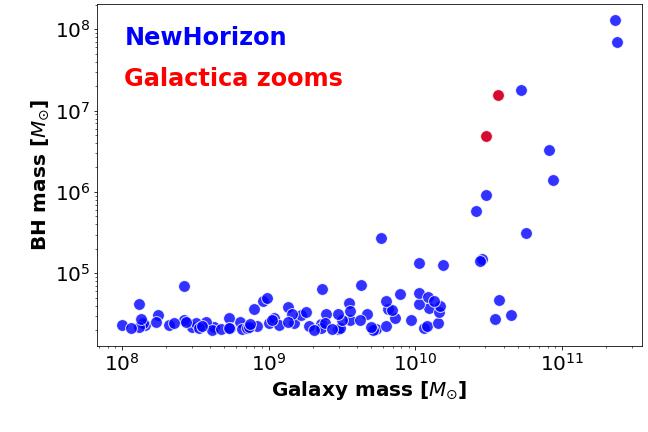}}
\caption{Variations of the primary BH mass with respect to their host
(stellar) galaxy mass at $z\sim0.18$. Our final catalog consists of 100 BHs
all in radio mode, including BHs extracted from the two \galacticas zooms.
Overall, across all stellar mass bins, BHs in \nhs are
under-massive in comparison to observations, lying up to one orders of magnitude below the bulk of the observed BH masses
\citep[see, for instance, Figure~22 of][]{nhz}.
}
\label{fig1}
\end{center}
 \end{figure}

The galaxy-BH catalog is produced using the same methodology described in
\cite{peirani+24} or adopted in
previous studies using either \hagn\, or \nh\,
\citep[e.g.,][]{volonteri+16,smethurst+23,beckmann+23}. 
At a given redshift, we link the primary BH to their host galaxy by
selecting the most massive BH to be contained within two half mass radii (hereafter R$_{1/2}$) of the galaxy's center, using an iterative loop. The other BHs contained within two half mass radii are labeled as "secondary" or "wandering" BHs and are discarded from our analysis.
In this scheme, we use a shrinking sphere approach \citep{power+03} to
determine precisely the galaxy center.
The half-mass radius\footnote{In \cite{peirani+24}, R$_{1/2}$ is called the effective radius.
In the present analysis, we distinguish between the half-mass radius
R$_{1/2}$, computed from the distribution of star particles, and
the half-light radius or effective radius \Reff, estimated from each synthetic
optical image (see section~\ref{subsec:opt}).}  
of each galaxy is estimated by taking the geometric mean of the
half-mass radius of the projected stellar densities along each of the
simulation's Cartesian axes. 
Note that BHs with mass below 2$\times$10$^4$M$_\odot$
are discarded from the analysis because they are too close
to the initial seed BH mass and likely to suffer from mass
resolution.

Since \nh\, is a zoom simulation, 
  low-mass resolution dark matter particles might ``pollute'' some halos,
  especially when they are located close to the boundary of the high
  resolution area. We however allow the selection of 
BHs in ``contaminated'' DM halos if the low resolution DM particles represent less than 0.1\% of the total mass of the halos.

We also only select galaxies with a stellar mass\footnote{returned by the  {\mbox{{\sc \small Adaptahop}}} structure finder \citep{aubertetal04, tweed+09}.} 
greater than 10$^8$M$_\odot$ to be more consistent
with galaxies selected in the MaNGA survey \citep[Mapping Nearby Galaxies at Apache Point Observatory,][]{manga}.
All these constraints lead to a sample of 100 galaxy-BH pairs.
We have checked that the 
different BHs are in radio mode. 
If a given BH is not in radio mode at the last
snapshot ($z=0.18$), we select the closest snapshot (that is, at a slightly higher redshift) where this constraint is satisfied.

Fig.~\ref{fig1} shows the variation of primary BHs mass against their host galaxy
mass at redshift $\sim$0.18.
As previously noted by \cite{nhz}, central massive black holes in \nhs\, typically experience significant growth only in galaxies with stellar masses exceeding a few 
$10^{10}$ M$_\odot$. In contrast, for galaxies with stellar masses below 5$\times$10$^9$ M$_\odot$, BH growth is generally suppressed by supernova feedback
\cite[see also][]{dubois+15, habouzit+17,trebitschetal17, lapiner+21}.
Additionally, because of their low initial (seed) masses, BHs in these low-mass systems (particularly dwarf galaxies) often struggle to remain anchored at the centers of their host galaxies, as highlighted by \cite{beckmann+23a_pop}. As a result, the majority of BHs in the simulation undergo minimal growth throughout its duration.
Overall, across all stellar mass bins, BHs in our sample are
slightly under-massive compare to observations, lying up to one order of magnitude below the bulk of the observed BH masses \citep[see, for instance, Figure~22 of][]{nhz}.
Yet our model reproduces both the slope of the local black hole mass-total stellar mass relation and the break that appears to be emerging from observational constraints
\citep[e.g.,][]{reines+15}. In our framework, BH growth remains inefficient in dwarf galaxies, becomes more effective once the host reaches sufficient mass ($\sim$3$\times$10$^9$ M$_\odot$), and eventually enters an AGN self-regulated phase. This mass-dependent transition captures the key qualitative features implied by current observations and theoretical models of BH-galaxy co-evolution.

\subsection{Synthetic DESI-LS and Euclid Optical Images}
\label{subsec:opt}

To facilitate comparison with recent observational trends, particularly those reported by 
\cite{zheng+24}, \cite{fernandez+25} and \cite{jung+25}, we produced a set of synthetic DESI-LS \citep{desi} like $r$-band optical images. For this purpose, we employed the \sunset\, code, a module included in the \ramses\, package that generates realistic galaxy photometry, images, and spectra.
Each stellar particle is assumed to behave as a single stellar population. We use here a Chabrier initial mass function \citep{chabrier03}, and we adopt 
the stellar population synthesis model of \cite{bruzual&charlot03} to compute the contribution of each stellar particle to the mock image.
Although the impact of dust attenuation is expected to be limited in the $r$-band, the original version of the code was updated to incorporate dust effects. 
To this regard, the dust column density in front of each stellar particle is computed using the gas-phase metallicity as a proxy for the dust distribution. The contribution of all gas cells in front of a particle is accounted for, assuming that 40\% of the mass of metals in gas cells are locked in dust grains which is the fraction in the Milky-Way \citep[see][]{dwek1998} and in many other evolved galaxies \citep{remy-ruyer+14}. To compute attenuation by dust, we adopt the value of the ratio of visual extinction to reddening $R_V = 3.1$ for the Milky Way dust grain model presented by \cite{Weingartner01}. 
While more advanced radiative transfer codes such as \skirt\, \citep{skirt1,skirt2} are available for simulating photon propagation in astrophysical systems, \sunset\, offers a practical compromise between physical realism and computational efficiency.

Following the methodology applied in \cite{peirani+24} for analyzing projected 2-d orientation angles, we used a Monte Carlo approach to generate the synthetic images. For each of the 100 galaxies in our sample, we use \sunset\, to create 50 synthetic optical images by randomly assigning a redshift in the range $0.02 < z < 0.18$ and a random spatial orientation. These mock images, generated by "observing" the original galaxy from different angles and redshifts, are diverse enough to be considered as statistically independent realizations. Each image was produced in the $r$-filter at DESI-LS resolution (0.262 arcsec/pixel) and convolved with a point-spread function (PSF) characterized by a full width at half maximum (FWHM) of 1.2 arcsec \citep{desi}. No further corrections, such as seeing adjustments, image noise and sky background were applied.

In parallel, we have generated a second set of mock observations using the specifications of the Euclid mission \citep{euclid_overview}: a resolution of 0.1 arcsec/pixel, 
the $I_{\rm E}$-band (or RIZ band), and a PSF with a FWHM of 0.16 arcsec \citep{mccracken+2025}. For consistency, we used the same set of galaxies, orientations, and redshift assignments as those employed for the DESI-LS-like images.

\subsection{Synthetic MaNGA velocity fields}
\label{subsec:vel}

The MaNGA survey \citep{manga} employs integral field units (IFUs) composed of fiber arrays arranged in a hexagonal pattern. There are five IFU configurations consisting of 19, 37, 61, 91, and 127 fibers, corresponding to diameters of $12.5^{\prime\prime}$, $17.5^{\prime\prime}$, $22.5^{\prime\prime}$, $27.5^{\prime\prime}$, and $32.5^{\prime\prime}$, respectively \citep{drory+15}. The galaxy sample is divided into two subsets: the Primary and Secondary samples, which are optimized to achieve spatial coverage out to $1.5\,R_\mathrm{e}$ and $2.5\,R_\mathrm{e}$, respectively.

To generate synthetic stellar velocity fields comparable to MaNGA observations, we first assign an IFU to each of the 5,000 galaxies in our mock sample. For this, we estimate the effective radius of each galaxy using DESI-LS-like synthetic $r$-band images, assuming a fixed redshift of $z = 0.037$ (i.e., the median redshift of the MaNGA survey). The galaxies are then randomly divided into two groups that reproduce the observed proportions of the Primary and Secondary samples. Each galaxy is subsequently assigned a specific IFU configuration to ensure optimal spatial coverage.

Next, we begin by computing the projected, light-weighted velocity fields in the $r$-band at DESI-LS resolution, convolved with a PSF of FWHM $=1.5$ arcsec 
as a simplified emulation of the typical atmospheric seeing at the telescope site
\citep{yang+16,bottrell+22,sarmiento+23}.
From these velocity maps, we calculate the mean velocity in each IFU fiber by averaging over all pixels located within the corresponding fiber aperture. The target galaxy is then "observed" through the assigned IFU using three dithering positions \citep[see, for instance, Fig.~4 of ][]{sarmiento+23}.
Finally, the fiber values are 
recombined and sampled onto a grid of  0.5$\times$0.5 arcsec$^2$ pixels.
We follow here the same procedure as described in \cite{law+16}, namely
the contribution of  a fiber $i$ to a specific pixel "p" is
given by:
\begin{equation}
    w_{p,i} = \frac{e^{-\frac{r_{p,i}^2}{2\sigma^2}}}{w_T},
\end{equation}
\noindent 
where $r_{p,i}$ is the distance between the pixel "p" and the center of the fiber "i", $\sigma=0.7\arcsec$ \citep{law+16} and $w_T$
a normalizing factor to keep the flux constant.

For other sophisticated methods for generating MaNGA images from cosmological 
simulations, we refer the readers to \cite{bottrell+22}, \cite{nanni+22}
and \cite{sarmiento+23}
as well as \cite{barrientos+23} for SAMI-like synthetic observations.

\subsection{Measurement of ellipticity and effective radius}
\label{sec:ell_Re}

For each of the 5,000 DESI-LS or Euclid synthetic optical images, 
the semi-major and semi-minor axis lengths, ellipticity and orientation
angle $\phi$ (angle between the semi-major axis and the $x$-axis) are computed 
on the basis of the inertia tensor $I_{\alpha\beta}$ of the 2-d flux
distribution: 

\begin{equation}
I_{\alpha\beta} = \sum_{i=1}^{N} L^{(i)}x_\alpha^{(i)}x_\beta^{(i)}\,\,\,\,\,\,\,\,\,\,\,\,\,\, (\alpha,\beta=1,2), 
\end{equation}

\begin{equation}
\mathrm{cov(I)} = 
\begin{pmatrix} I_{11} & I_{12}\\ I_{12} & I_{22} \end{pmatrix},
\end{equation}

\begin{equation}
\phi = \frac{1}{2} \arctan \left( \frac{2I_{12}}{I_{11} - I_{22}} \right),
\label{eq:oa}
\end{equation}

\noindent
where $L^{(i)}$ and $x_\alpha^{(i)}$ are the flux and the projected position vector of the $i$-th pixel within a given
enclosed flux or light region (specified by the value of $N$).
That intertia tensor is then diagonalized and the square root of the eigenvalues
give the relative size of the semi-major and minor axis. The orientation angle can be
derived from the eigenvalue vectors and its expression is summarized
in Equ.~\ref{eq:oa}.
Note that a similar methodology have been used in other theoretical studies
using large cosmological simulations \citep[see, for instance,][]{suto+17,okabe+18, lagos+18,rodriguez-gomez+19,araujo-ferreira+25}.

\begin{figure}
\begin{center}
\rotatebox{0}{\includegraphics[width=\columnwidth]{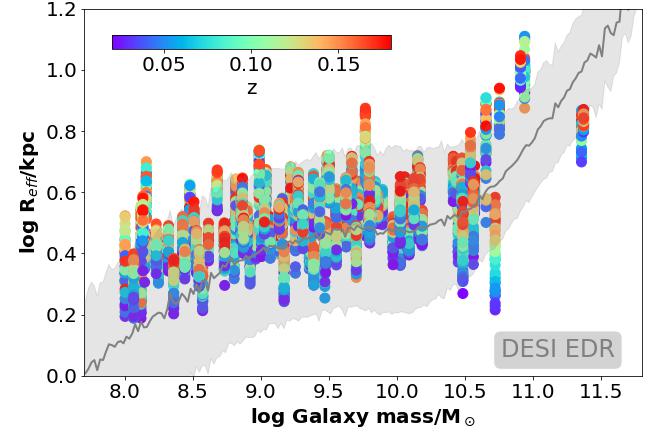}}
\rotatebox{0}{\includegraphics[width=\columnwidth]{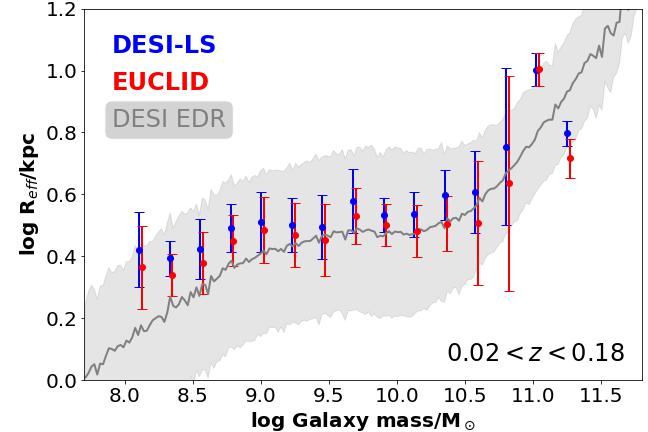}}
\caption{
Variations of the effective radius estimated from synthetic DESI-LS $r$-band images
with respect to the galaxy mass (upper panel). For each galaxy, we generated 50 synthetic images by assigning random spatial orientations and redshifts in the range $0.02 < z < 0.18$. For qualitative comparison with observational data, the black line represents the average half-light radius measured from the DESI Early Data Release (EDR), based on a sample of 151,530 galaxies within the same redshift range. The grey shaded area denotes the dispersion around the mean value. Overall, the trend observed in our synthetic sample agrees
reasonably well with observational expectations, suggesting that
the theoretical estimates of \Reff\, are reliable.
In the lower panel, we compare the corresponding \Reff\, values derived from the synthetic Euclid images. On average, the effective radii estimated from the DESI-LS synthetic images are slightly larger than those obtained from Euclid, primarily due to the lower angular resolution of DESI-LS.
}
\label{fig_Re}
\end{center}
 \end{figure}

In the following, the effective radius \Reff\, is defined as half the length of the semi-major axis of the ellipse that encloses half of the total flux in the optical image. In Fig.~\ref{fig_Re}, we present the variation of the effective radius computed from the 5,000 synthetic DESI-LS $r$-band images as a function of stellar mass. For qualitative comparison with observational trends, we also show the mean and dispersion of half-light radii derived from the DESI Early Data Release \citep{desi_ED}, selecting all galaxies within the redshift range $0.02 < z < 0.18$.

Overall, the trend observed in our synthetic sample agrees reasonably well with observational expectations, suggesting that the theoretical estimates of \Reff\, are acceptable. However, we note a slight systematic excess in \Reff\, for low-mass galaxies ($M_\star < 10^9\,M_\odot$) compared to observations. This discrepancy may arise from limitations in image resolution and/or the treatment of feedback processes in the simulation \citep{martin+25,watkins+25}.

In fact, the estimation of \Reff\, appears to be resolution-dependent. When the number of pixels covering a galaxy is reduced (such as at higher redshifts) the effective radius tends to be overestimated. This effect is illustrated in the lower panel of Fig.~\ref{fig_Re}, where we compare the mean \Reff\, values obtained from DESI-LS and Euclid synthetic images. Although the filters and PSF convolution kernels differ between the two surveys, it is evident that the DESI-LS based estimates of $R_\mathrm{e}$ are, on average, systematically larger than those derived from Euclid. 
This resolution bias is also noticeable in the upper panel of Fig.~\ref{fig_Re} suggesting that, for a particularly compact object, the estimated effective radius increases with redshift, corresponding to a decrease in pixel angular resolution.

\subsection{Optical and kinematic position angles estimation}

Throughout this work, position angles are defined with respect to the $x$-axis (i.e.,the horizontal axis) and range from $0^\circ$ to $180^\circ$.

We define the optical position angle (\PAo) for each DESI-LS and Euclid synthetic image as the angle between the semi-minor axis of the projected ellipse (where the semi-major axis corresponds to the effective radius) and the $x$-axis, following the methodology described in Section~\ref{sec:ell_Re}. We also compare trends measured at 2\Reff.
Note that in the DESI Legacy Survey, the optical position angle, \PAo, is not measured explicitly at the effective radius but is inferred from a global model fit to the galaxy's surface brightness profile. Since the fit is most constrained around the effective radius (where the signal-to-noise is good), the PA effectively reflects the orientation around \Reff.

To estimate the kinematic position angles, \PAk, from each synthetic MaNGA velocity field,
we use the \mbox{{\sc \small PaFit}} package \citep{krajnovic+06}, a Python-based tool 
to determine the global kinematic position angle of galaxies (with 3\sigPA values)
and widely used in MaNGA observational analysis.
The \mbox{{\sc \small PaFit}} routine basically performs a symmetry-based fitting of the observed velocity field and then tries different trial position angles and centers to find the configuration that minimizes the difference between the observed velocity field and the symmetrized model.

\begin{figure*}
\begin{center}
\rotatebox{0}{\includegraphics[width=18.5cm]{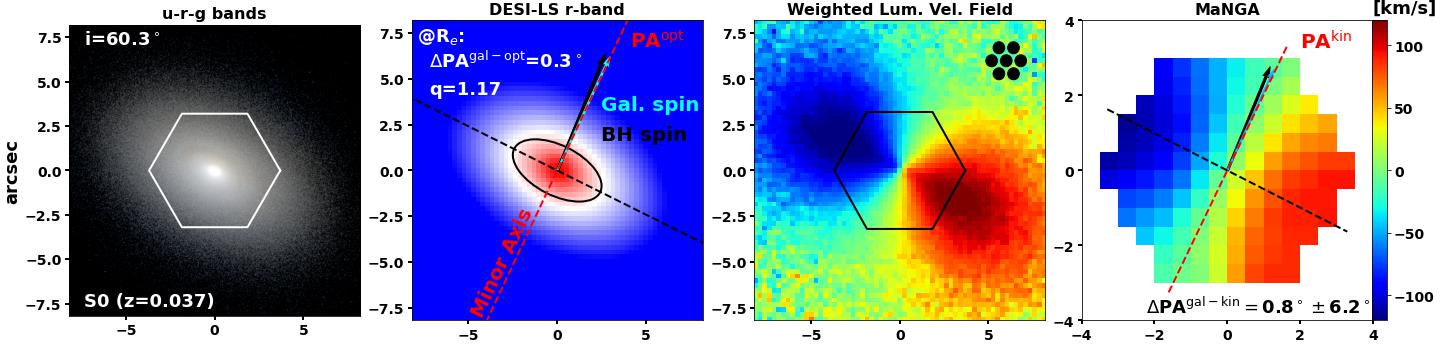}}
\rotatebox{0}{\includegraphics[width=18.5cm]{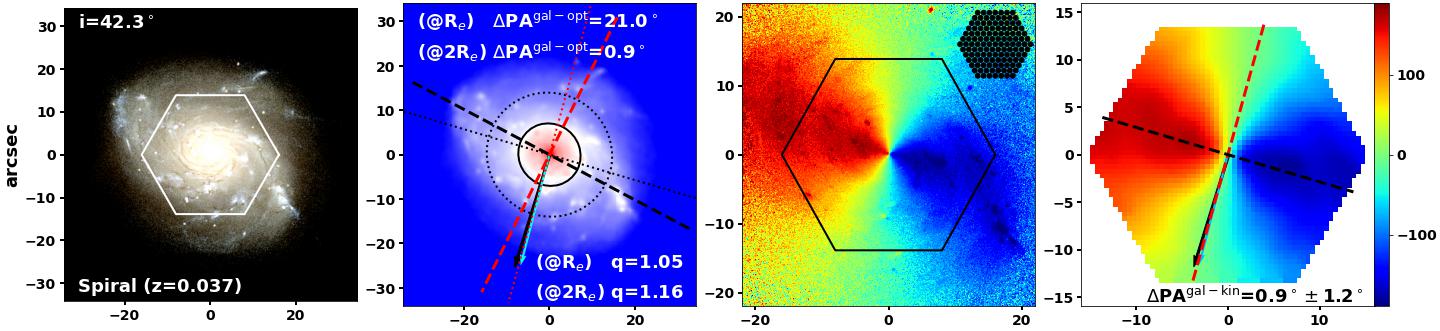}}
\rotatebox{0}{\includegraphics[width=18.5cm]{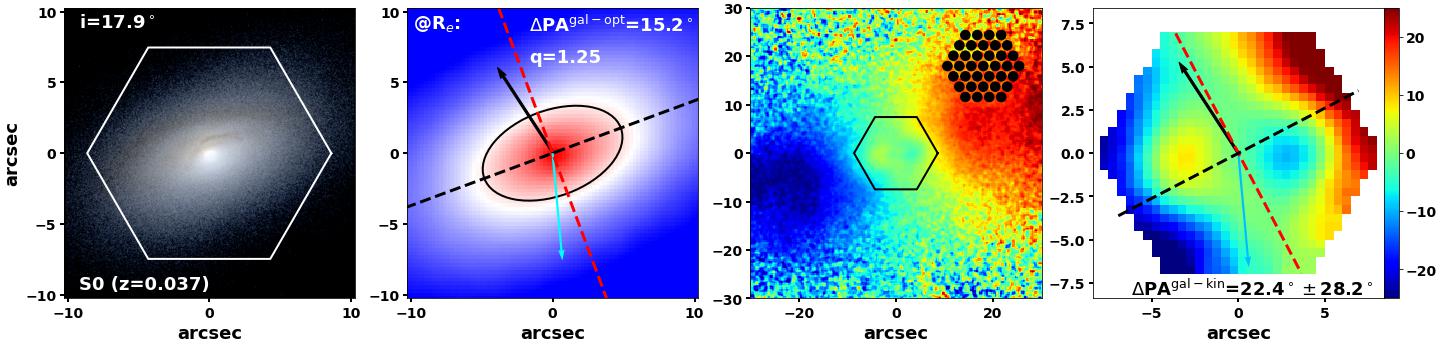}}
\caption{Examples of synthetic optical and kinematic images derived
in our analysis. Each row corresponds to a specific galaxy.
The first column shows high resolution images using $u$-$g$-$r$ bands, along with the
value of the inclination angle ($i$) at which the galactic plan
is viewed by the observer \citep[see Fig.~19 in][]{peirani+24}.
The second column presents synthetic DESI-LS-like $r$-band images with a resolution of 0.262 arcsec/pixel, convolved with a PSF having a FWHM of 1.2 arcsec. The black and red dashed lines indicate the orientations of  the major and minor photometric axes, respectively. The semi-major axis of each ellipse corresponds to the effective radius (\Reff), while the orientation of the semi-minor axis serves as a proxy for the direction of the projected stellar angular momentum. For comparison, these panels also include the projected BH spin vector (in black) and the projected stellar angular momentum vector (in cyan), the latter estimated from all stars within the half-mass radius.
In the second galaxy, the dotted line and the dotted ellipse indicate measurements at 2\Reff. The axis ratio $q$ denotes the ratio of the semi-major to semi-minor axes.
In the third column, we show the corresponding high-resolution projected velocity fields, while the fourth column displays the synthetic MaNGA-like velocity fields. These latter are computed on a 0.5 $\times$ 0.5 arcsec/pixel grid, based on the interpolation of three dithering fiber configuration illustrated in the top-right corner of the third-column panels. Kinematic position angles (\PAk), derived using the \mbox{{\sc \small PaFit}} package \citep{krajnovic+06}, are indicated with red dashed lines.
In the first example, both the optical and kinematic PAs closely trace the direction of the projected stellar spin. In the second example, the optical PA is less reliable due to the near-circular shape of the galaxy at one effective radius ($q \approx 1$), while the kinematic PA more accurately captures the spin direction. The third example features a spheroidal galaxy with a kinematically decoupled core (a central stellar component rotating counter to the outer stellar population). This complex structure is not evident from the optical morphology alone, resulting in a more noticeable misalignment of \dPAgalopt=15.2$^\circ$. The kinematic estimate performs even worse, with \dPAgalkin=22.4$^\circ$ and a substantial 3\sigPA uncertainty of $28.2^\circ$, likely due to the presence of the decoupled core. 
}
\label{fig_examples}
\end{center}
 \end{figure*}

Following the definition in \cite{franx+91}, one can now estimate the
misalignment angles \dPAXY\, as the difference between the position angle
\PAX\,  and the position angle \PAY\, as:

\begin{equation}
    \mathrm{sin}\,( \Delta \mathrm{PA^{\mathrm{X-Y}}} ) = \lvert\,\mathrm{sin(\,PA}^{\mathrm{X}}- \mathrm{PA}^{\mathrm{Y}}\,)\,\rvert.
\end{equation}

In this parameterization,
\dPAXY\,
 lies between 0\degrees\, and 90\degrees\, and is not sensitive to differences of
180\degrees\, between \PAX\, and \PAY.

To illustrate how accurately the optical and kinematic position angles trace the orientation of the "true" or intrinsic projected angular momentum of a galaxy, we present in Fig.~\ref{fig_examples} three representative examples from our sample. For each galaxy, we compute the misalignment angles \dPAgalopt\, and \dPAgalkin\, defined respectively as the angular differences between the projected galaxy spin \PAgal\, and the optical (\PAo) or kinematic (\PAk) position angles.
The first galaxy is an S0 with a passive evolutionary history. The second is a typical spiral galaxy, while the third is another S0 galaxy featuring a kinematically decoupled core \citep[see, for instance,][]{peirani+25}. The first column of Fig.~\ref{fig_examples} displays $u$–$g$–$r$ composite images of these galaxies to provide a clear view of their projected morphological shapes.
The second column shows the corresponding synthetic DESI-LS $r$-band images, convolved with an appropriate PSF. Each overlaid ellipse (solid line) encloses half of the galaxy's total light. The semi-major axis of the ellipse defines the effective radius, while the orientation of the semi-minor axis serves as a proxy for the direction of the projected stellar angular momentum.
For comparison, these panels also include the projected BH spin vector (in black) and the projected stellar angular momentum vector (in cyan), the latter estimated from all stars within the half-mass radius (as defined above).
The third column presents the light-weighted velocity fields at a high resolution. The fourth column shows the corresponding MaNGA-like velocity fields, along with the inferred kinematic position angle \PAk\, (red dashed line; see section~\ref{subsec:vel}).

In the first case (top row), the galaxy is an S0 with a passive evolutionary history and a relatively smooth light distribution, although some dust attenuation features are visible. The projected stellar angular momentum vector is nearly aligned with the semi-minor axis, yielding a misalignment angle of 
\dPAgalopt=0.3\degrees.
For the kinematic analysis, this galaxy is observed out to 1.5\Reff\, using a MaNGA IFU with 7 fibers, due to its relatively compact size (\Reff$=$1.97 kpc). The position angle determined from \mbox{{\sc \small PaFit}} fitting routine gives a misalignment of \dPAgalkin=$0.8^\circ \pm 6.2^\circ$, also indicating a very good agreement with the projected galaxy spin direction.

In the second case, the ratio of the semi-major to semi-minor axis measured at
\Reff\, is close to unity ($q=1.05$). As a result, the orientation of the semi-minor axis becomes less reliable for estimating the direction of the true projected stellar angular momentum. In this configuration, we find a misalignment of \dPAgalopt=11.0$^\circ$ which
is still low. However, the accuracy improves significantly when the position angle is measured at 2\Reff\,(that is, \dPAgalopt=0.9$^\circ$, where $q=1.16$).
In contrast, the velocity field exhibits the typical signature of a spiral galaxy with strong rotational support. The synthetic MaNGA velocity map yields a kinematic misalignment of \dPAgalkin=0.9$^\circ$, with a high degree of confidence (3\sigPA uncertainty of $1.2^\circ$).

The third case features an atypical spheroidal galaxy with a kinematically decoupled core, corresponding to a central stellar population that rotates in the opposite direction to the outer stellar component, a feature clearly visible in the synthetic velocity field.
In this particular case, the main galaxy has undergone a merger with another (satellite) galaxy, with a mass ratio of 1:4. The stellar population that originally belonged to the satellite galaxy is now found at larger radii following the merger event and exhibits a counter-rotation with respect to the main stellar component. 
The gas accreted from the satellite galaxy has completely replaced the original gas of the host and also counter-rotates relative to the main stellar component.
Consequently, as indicated by the stellar velocity field (third row, right panel), the BH spin, which aligns with the orientation of the accreted gas, follows the angular momentum of the outer stellar population rather than that of the central, decoupled core, which might have been expected initially.
For more details on the formation of this particular galaxy, as well as a broader theoretical framework on the origin of counter-rotating galaxies, we refer the reader to galaxy G-31 in \cite{peirani+25}, and to \cite{peirani+24} for the consequences on the evolution of its central black hole, BH-549. 
In this example, the optical morphology alone does not reveal the presence of such a complex kinematic structure. However, the alignment is less evident, with a value of  \dPAgalopt=15.2$^\circ$. The kinematic prediction performs even worse, with \dPAgalkin=22.4$^\circ$ and a large 3\sigPA uncertainty of $28.2^\circ$, likely due to the presence of the decoupled core.
Note that, interestingly,  the synthetic MaNGA image derived for this galaxy
looks quite similar to the stellar MaNGA velocicy field derived from the galaxy PGC 66551
and studied in \cite{katkov+24}.

\section{Results}
\label{sec:results}

\subsection{Galaxy spin - Position angles misalignment}
\label{subsec:opt-kin-mis}

\begin{figure*}
\begin{center}
\rotatebox{0}{\includegraphics[width=18cm]{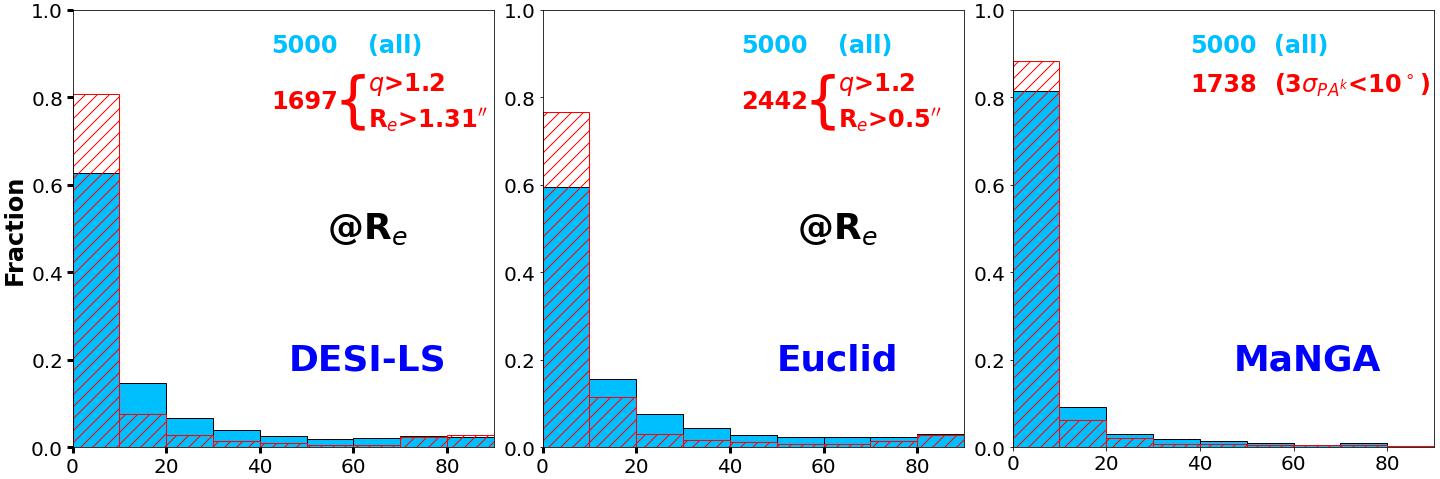}}
\rotatebox{0}{\includegraphics[width=18cm]{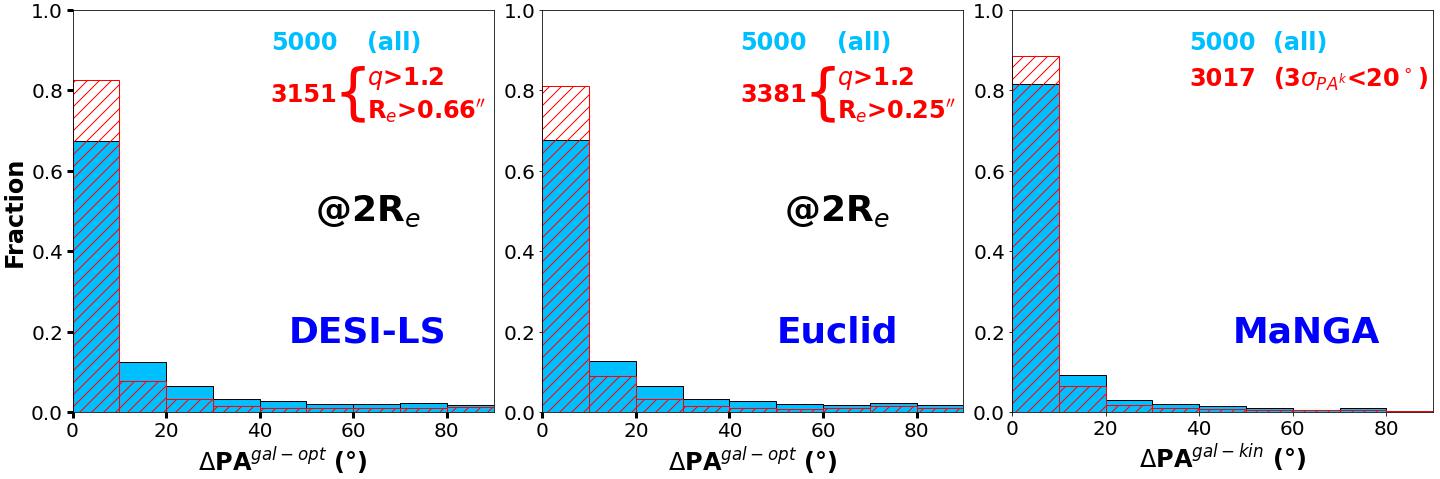}}
\caption{Distributions of \dPAgalopt\, and \dPAgalkin.
This figure presents the distributions of \dPAgalopt, the 2-d misalignment angle between the projected stellar angular momentum vector of the galaxy and the optical position angle, defined as the orientation of the minor axis of the projected light ellipse. The first and second columns correspond to synthetic DESI-LS and Euclid images, respectively, with optical PAs estimated at one effective radius (top row) and at two effective radii (middle row). The third column shows the distributions of \dPAgalkin, defined as the angle between the projected spin vector of the galaxy and the kinematic PA derived from synthetic MaNGA-like stellar velocity fields.
In all panels, red hatched histograms indicate results when only "good cases"
(i.e., more reliable PA measurements) are selected.
We also specify the number of "sources" or images used to derive
the different histograms.
}
\label{fig_DPA_gal_pa}
\end{center}
 \end{figure*}

The first step of our analysis is to assess the reliability of optical and kinematic position angles in tracing the orientation of a galaxy’s projected angular momentum. For consistency, we adopt the same definition of galaxy angular momentum as in our previous studies \citep{peirani+24,peirani+25}, which is computed using all star particles located within the half-mass radius (as defined in Section~\ref{subsec:catalog}).

In Fig.~\ref{fig_DPA_gal_pa}, we show the distributions of \dPAgalopt, derived from our catalogs of 5,000 synthetic DESI-LS and Euclid optical images, and \dPAgalkin, derived from (5,000) MaNGA-like synthetic velocity fields. For comparison, the optical PAs are measured at both one \Reff\, (top row) and two \Reff\, (bottom row).

Following the methodology adopted in \cite{zheng+24} and \cite{fernandez+25}, we also analyze trends using only "good cases" (that is, galaxies with more reliable PA estimates). In our analysis, "good cases" for optical images are defined as those with a projected axis ratio $q>1.2$ and a ratio of \Reff\, (or 2\Reff) to the instrument's pixel resolution greater than 5. For instance, this corresponds to \Reff>1.13$^{\prime\prime}$ or
\Reff>0.66$^{\prime\prime}$ for DESI-LS, depending on whether the PA is measured at one or two effective radii.
For the kinematic maps, "good cases" are selected based on PA uncertainties (3\sigPA) being less than 10\degrees\, or 20\degrees\,, respectively. These thresholds were chosen to roughly match the number of "good cases" of DESI-LS optical images for each \Reff-based measurement.

Figure~\ref{fig_DPA_gal_pa} suggests that the kinematic PAs are a more accurate estimator for the "true" projected galaxy angular momentum. Approximately 80\% of galaxies exhibit \dPAgalkin<10\degrees\, while ~60\% meet this threshold for the optical PAs. If we focus on the "good cases", however, the three 
estimators perform almost equally well as 88.4\%, 88.0\%, and 94.4\% of the 
DESI-LS, Euclid, and MaNGA sub-samples, respectively, show 
$\Delta$PA<20\degrees.

We also find that estimating optical PAs  at 2\Reff\, slightly improves the overall alignment with the projected galaxy spin  \PAgal\, compared to measurements taken at \Reff, particularly for $\Delta$PA<10\degrees. While the overall trends remain similar to those obtained using PAs measured at one effective radius, the differences between the three approaches become less pronounced when focusing on "good cases". This suggests that, in well-resolved galaxies with reliable PA estimates, the dominant ``noise''
  source in estimating the true misalignment is likely due to projection effects rather than instrumental limitations such as point spread function width or filter selection.
Additionally, we observe no significant change in the distribution of
\dPAgalkin\, when applying a more stringent selection criterion of
3\sigPA<10$^\circ$, further supporting the robustness of the kinematic alignment trends.

We did not find significant improvement in alignment
statistics when switching from DESI-LS to Euclid, despite the latter’s
higher resolution. 
One possible explanation is that the
 simulated galaxies studied in this work are in general well resolved
 galaxies, even in DESI-LS. Euclid's larger sample size will however
 allow to extend this study at higher redshift/lower mass galaxies
  and will likely enable more precise conclusions in the future.

\begin{figure}
\begin{center}
\rotatebox{0}{\includegraphics[width=\columnwidth]{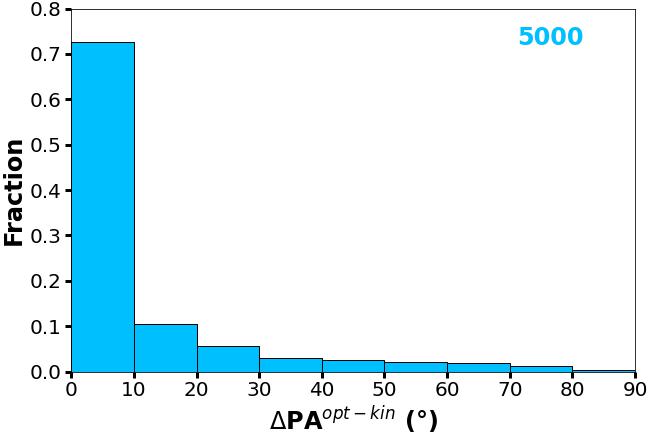}}
\caption{Histogram of \dPAoptkin, the angles between optical and kinematic PAs derived
from our sample of 5,000 synthetic DESI-LS r-band and MaNGA velocity fields.
72.8\% and 82.4\% of the sample have optical-kinematic misalignment lower
than 10\degrees\, and 20\degrees, respectively.
Optical PAs are estimated at two effective radii.
}
\label{fig_DPA_opt_kin}
\end{center}
 \end{figure}

Our findings are consistent with previous studies. First, the distributions of optical position angles shown in Fig.~\ref{fig_DPA_gal_pa} are similar to those obtained for elliptical galaxies from the EAGLE simulation \citep{fernandez+25}, which also exhibit a high fraction of cases with $\Delta$PA<20\degrees. Second, observational analyses generally report good agreement between optical and kinematic PAs. For example, based on a sample of approximately 2,300 MaNGA galaxies, \cite{graham+18} found a strong peak in the distribution for \dPAoptkin<30\degrees (see their Fig. 12). Specifically, they reported that 83.7\% of regular early-type galaxies (ETGs) and 84.5\% of spiral galaxies are aligned within \dPAoptkin<10\degrees, in good agreement with \cite{krajnovic+11}, who found that 90\% of 260 early-type galaxies from the ATLAS$^{3D}$ survey lie within \dPAoptkin<15$^\circ$.
Additionally, \cite{barrera-ballesteros+15} found that the morpho-kinematic misalignment is less than 22\degrees\, in 90\% of a sample of 103 interacting galaxies from the CALIFA survey (see their Fig. 4). In our case, the alignment trend is slightly weaker: as shown in Fig.~\ref{fig_DPA_opt_kin}, we find that 72.8\% and 82.4\% of the sample exhibit optical–kinematic misalignment below 10\degrees\, and 20\degrees, respectively.

\subsection{BH spin/Jet VS. PA misalignment}
\label{subsec:jet-mis}

\begin{figure*}
\begin{center}
\rotatebox{0}{\includegraphics[width=18cm]{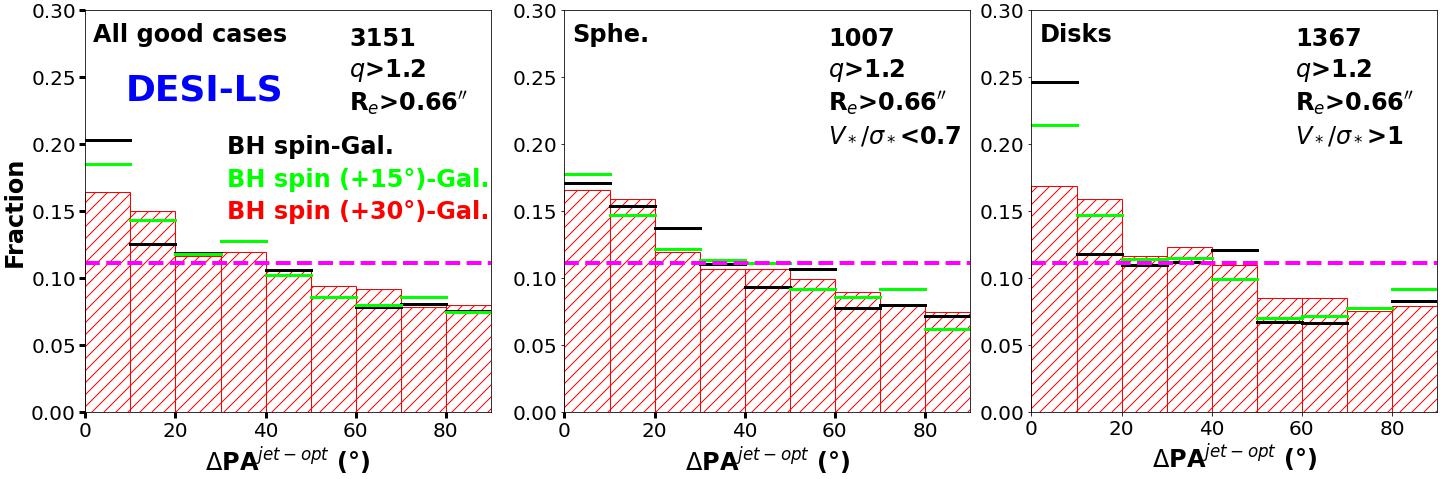}}
\rotatebox{0}{\includegraphics[width=18cm]{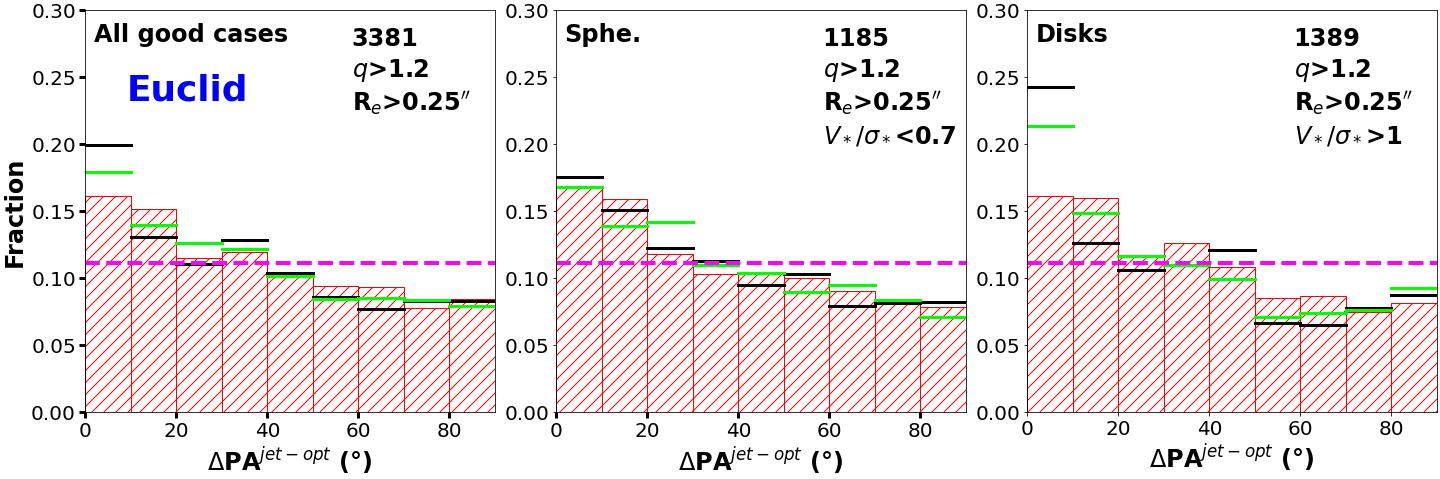}}
\rotatebox{0}{\includegraphics[width=18cm]{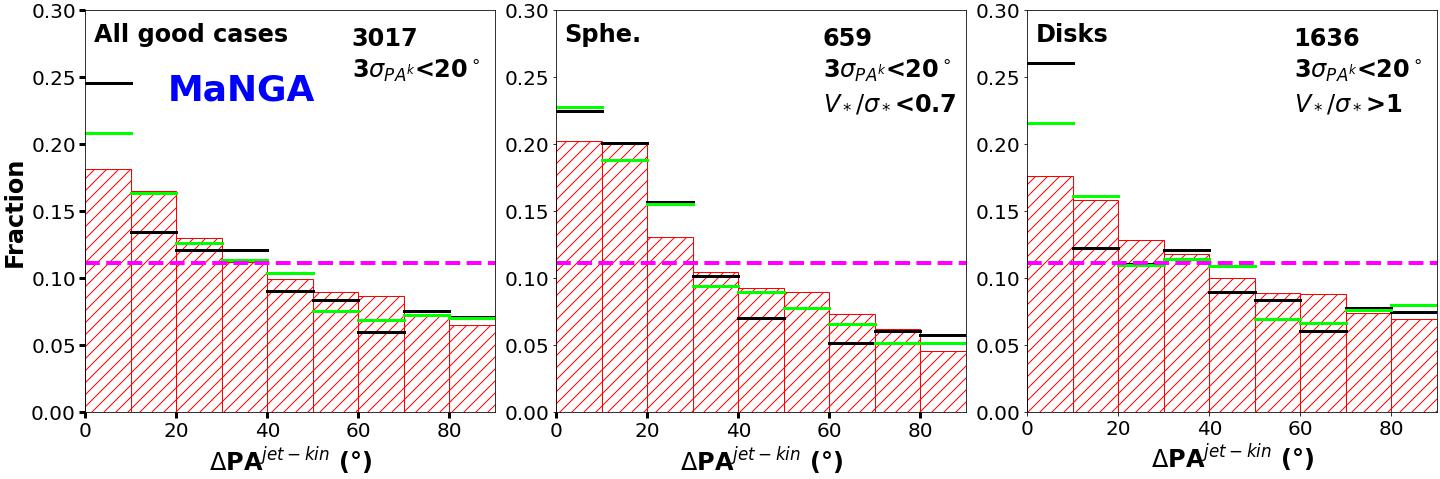}}
\caption{Histograms of \dPAjetopt, the two-dimensional misalignment angle between the projected galaxy spin and the optical position angle,  derived from synthetic images from the DESI-LS (top row) and Euclid (middle row) catalogs, with measurements estimated at 2\Reff. The third row presents histograms of \dPAjetkin, the misalignment between the projected galaxy spin and the kinematic PA derived from synthetic MaNGA velocity fields.
The first column includes results for the treatment of all "good cases", defined by 
expected more reliable optical or kinematic PA measurements,
while the second and third columns display trends for spheroid-dominated galaxies (\Vsig<0.7) and disk-dominated galaxies (\Vsig>1.0), respectively.
In each panel, the black, green, and red hashed histograms represent the results with no added scatter, and with uniform scatter of up to 15\degrees\, and 30\degrees\, respectively, applied to the 3-d BH spin vectors.
The dashed magenta line refers to the uniform distribution
and the number indicate the number of images used to derive
the histrograms.
Overall, a clear trend of alignment between the jet direction and the galaxy orientation is observed, with a higher fraction of galaxies exhibiting misalignment angles below 20\degrees, even when random angular scatter is introduced to the BH spin.
}
\label{fig_dPA_jet}
\end{center}
 \end{figure*}

\begin{figure*}
\begin{center}
\rotatebox{0}{\includegraphics[width=18cm]{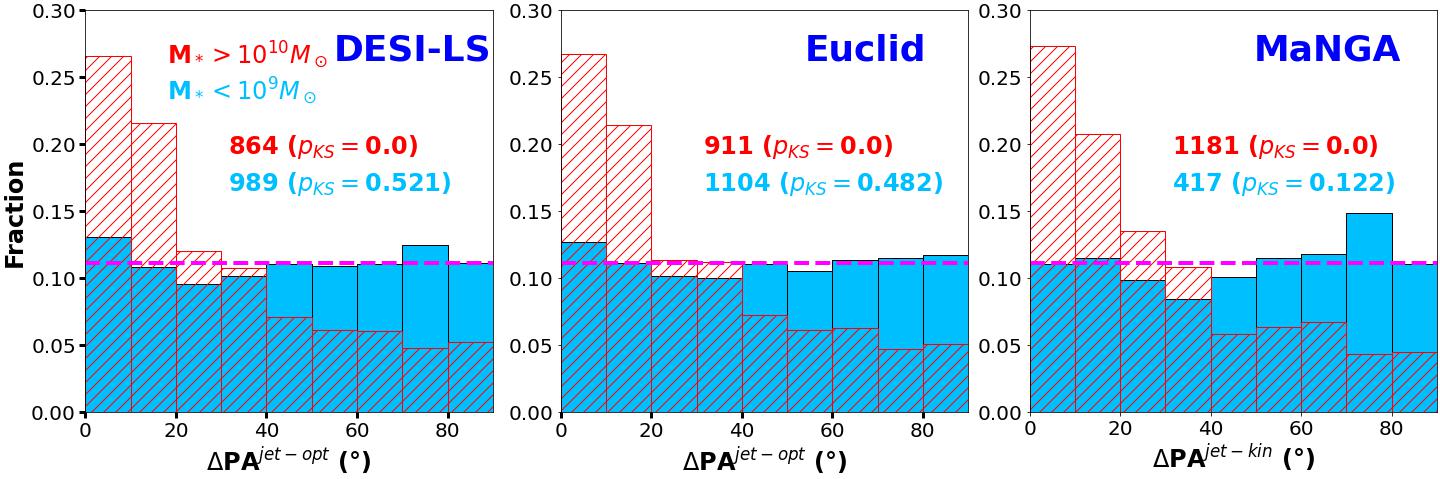}}
\caption{
Same as Fig.~\ref{fig_dPA_jet},  but here we distinguish between galaxies with stellar masses below 10$^9$M$_\odot$ and above 10$^{10}$M$_\odot$. The numbers indicated in each panel correspond to the sample sizes used to generate the histograms. Values in parentheses denote the Kolmogorov–Smirnov test $p_{KS}$-values. The results suggest that low-mass galaxies exhibit a more uniform distribution of BH–galaxy misalignment angles (characterized by $p_{KS}>0.05$). Note that only "good cases" are shown, assuming a uniform perturbation of the BH spin of 30\degrees\,, with no morphological distinction between spheroidal and disk galaxies.
}
\label{fig_dPA_minmax}
\end{center}
 \end{figure*}

We now turn to the core objective of this study: investigating the degree of misalignment between AGN jets and the orientation of their host galaxies. Although the \nhs simulation
does include AGN radio-mode feedback, clearly identifying jet structures in mock images remains challenging due to resolution limitations and the complexity of radio jet morphology.
Instead, since AGN jets are launched along the direction of the BH spin in the simulation, we use the BH spin vector as a proxy for jet orientation.
We note, however, that jets are intermittent phenomena and therefore not always observable, contrary to what is implicitly assumed in our approach. Incorporating a duty cycle for jet production and observability into our Monte Carlo framework is beyond the scope of the present paper, but should certainly be explored in future work to assess its potential impact. We also recall that all BHs in our sample are selected in the radio mode, which increases the likelihood of jet formation and thus partly alleviates concerns regarding duty-cycle realism.

It is important to acknowledge that AGN jets are not always perfectly aligned with the BH spin vector. As jets propagate from the accretion disk to galactic scales, their direction can be influenced by a number of factors:
interaction with the multiphase interstellar medium \citep[e.g.,][]{mukherjee+18, cielo+18, M87,borodina+25},
jet precession \citep{duun+06, krause+19,ubertosi+24},
or motion through the intracluster medium \citep[e.g.,][and references therein]{heinz+06,morsony+10,morsony+13,odea+23}.
This potential issue might be partly alleviated in recent observational analyses \citep{fernandez+25,zheng+24,jung+25}   
by using the Very-long-baseline Interferometry (VLBI) to see the immediate surrounding of the BHs.

Moreover, as discussed in section~\ref{sec:simu}, we stress again that the gas accretion disk around the black hole is not spatially resolved in our simulations. 
we indeed assumed that the angular momentum of the accretion disk aligns with that of
the gas measured at a distance of $\sim$136 pc (i.e., four times the highest spatial resolution) from each black hole. 
This approximation could become problematic in cases where the nuclear axis is misaligned with the larger-scale disk axis, as suggested by the high-resolution simulations of \cite{hopkins+12}.

Therefore, to account for potential misalignments between the nuclear axis and the large-scale galactic axis, as well as uncertainties related to jet propagation and projection effects, we introduce an artificial angular scatter to the 3-d BH spin orientations. Specifically, we apply a uniform random perturbation between 0\degrees\, and 30\degrees\, to mimic misalignments that may arise during gas accretion episodes and/or as the jet interacts with the multiphase ISM during its propagation to larger scales. 
Larger scatter amplitudes would progressively dilute any intrinsic alignment signal, eventually driving the distribution toward a uniform one and making any physical correlation much more difficult to detect. Conversely, smaller scatter values preserve tighter alignment signatures, but may underestimate the true level of physical and observational uncertainties. Our adopted range therefore represents a conservative compromise for exploring how such uncertainties affect the predicted misalignment distributions.
The jet position angle  is then determined by projecting this perturbed spin vector in 2-d, using the orientation of the host galaxy.

 Fig.~\ref{fig_dPA_jet} shows the distribution of \dPAjetopt, the angle between the projected jet direction and the optical PA, the latter derived from both DESI-LS and Euclid synthetic images\footnote{ Note that because AGN jets are not expected to be visible in Euclid images, future studies of radio/optical orientation will need to incorporate VLBI observations again.}.
 For clarity, we present results only for the “good cases” and consider three models: one with no added scatter, and two with uniform scatter of up to 15\degrees\, and 30\degrees\,, respectively, applied to the 3-d BH spin vectors.
Trends for disk-dominated galaxies and spheroid-dominated galaxies are shown in the second and third columns, respectively. To simplify the analysis, and rather than following a more observationally consistent approach, we classify galaxies using the ratio of stellar rotational velocity ($V_*$) to velocity dispersion ($\sigma_*$), derived from the simulations. We adopt \Vsig>1.0 for disk-dominated systems  and \Vsig<0.7 for spheroid-dominated systems \citep[values commonly used in previous studies e.g.,][]{dubois-seb16,peirani+17}
We also compute \dPAjetkin, the misalignment between jets and the kinematic PA derived from synthetic MaNGA-like velocity fields. 
Note that, for optical images, we only report results using PAs measured at
2\Reff, as trends at \Reff\, are similar.

Our results indicate a statistically significant
alignment trend even after taking account of  realistic observational errors. Indeed, a large fraction of galaxies display $\Delta$PA<20\degrees, consistent with a physical alignment between jets and galaxy orientation. 
 A modest perturbation of 15\degrees\, already reduces the strength of the signal, shifting the \dPAjetopt\, distribution closer to uniformity, although a clear excess of aligned systems remains. When the scatter is increased to 30\degrees, the dilution becomes more pronounced, as expected, yet the residual alignment persists at a statistically significant level.
This systematic weakening of the signal with increasing perturbation amplitudes highlights both the robustness of the underlying alignment and the sensitivity of projected \dPAjetopt\,\, measurements to small-scale physical processes that can alter jet propagation.
This confirms the findings of \cite{peirani+24}, where projected BH spin vectors were shown to align with the intrinsic galaxy spin.

Interestingly, we do not observe a strong difference in alignment trends between spheroidal (elliptical) and disk-dominated (spiral) galaxies. This is somewhat surprising
as spiral galaxies typically assemble via secular internal evolution
\citep[e.g.,][]{kormendy_and_kennicutt04} and the steady accretion of cold gas along cosmic filaments 
\citep[e.g.,][]{keres+05,dekel+06,dekel+09,davinocich+15},
whereas elliptical galaxies form mainly through major or repeated minor mergers that transform rotationally supported disks into pressure-supported spheroids
\citep[e.g.,][]{white78,barnes92,hernquist92,Naab+07,bournaud+11,rodriguez-gomez+16}.
These differing evolutionary paths are thus expected to influence the orientation and stability of the central black hole's spin axis.

Here also, the distributions from DESI-LS and Euclid mock images are similar, despite Euclid's superior spatial resolution. However, Euclid's broader sky coverage and deeper imaging are expected to improve sample statistics in future analyses.
Also, both optical and kinematic PAs yield comparable alignment trends, though the kinematic measurements show a slightly stronger alignment signal.

We have also performed a Kolmog\'orov–Smirnov (KS) test for each distribution of
Fig.~\ref{fig_dPA_jet} and analyzed the corresponding $p_{KS}$-values, which offer a binning-independent assessment of the distribution shape. The KS test is a non-parametric statistical method used to evaluate the goodness of fit between an empirical distribution and a reference distribution. In our case, we compare the observed distributions against a uniform distribution. In all cases presented
in Fig.~\ref{fig_dPA_jet}, the resulting $p_{KS}$-values tend toward zero, indicating that the observed distributions significantly deviate from uniformity.

Our findings agree well with recent VLBI-based observational studies. In particular,
\cite{fernandez+25} presented histograms of jet–galaxy PA differences using VLBI jets cross-matched with DESI Legacy Survey data (see their Fig. 3), showing similar alignment tendencies.
\cite{jung+25} reported that 39.8\% of "extended jet" sample have jets aligned within 30\degrees. Our results show slightly higher alignment rates:
 43.0\%, 44.4\%, and 42.0\% for all, elliptical, and disk-dominated galaxies, respectively (DESI-LS, "good cases"), and 47.3\% for the MaNGA-based analysis.

\cite{zheng+24} also studied radio–optical PA misalignments using data from LOFAR (LoTSS DR2), FIRST, DESI Legacy Surveys, and SDSS. They found a prominent minor-axis alignment trend for radio AGNs, possibly more pronounced than in our results. This could be due to their focus on more massive systems, where jet alignment may be more tightly correlated with host galaxy structure.

To conclude, we examine the impact of galaxy stellar mass on the distribution of jet–galaxy misalignment angles. As shown in \cite{peirani+24}, black holes  residing in low-mass galaxies generally exhibit spins that are less well-aligned with the angular momentum of their host galaxies compared to those in high-mass systems. This trend arises from the fact that, as discussed in section~\ref{subsec:catalog},  BHs in low-mass galaxies are often off-centered and experience limited growth due to inefficient gas accretion.
The results are presented in Fig.~\ref{fig_dPA_minmax}, where we distinguish between galaxies with stellar masses below 10$^9$M$_\odot$ and those above 10$^{10}$M$_\odot$.
Note that we didn't include intermediate-mass galaxies (between 10$^9$M$_\odot$ and  10$^{10}$M$_\odot$) to emphasize the contrast between the low- and high-mass regime trends.
Also, for clarity, we show only the "good cases" and assume a uniform perturbation of 30\degrees\, applied to the BH spin vectors. No morphological classification (e.g., disk or spheroid) is made in this analysis.
Our findings reveal that, in low-mass galaxies, the distribution of jet–galaxy misalignment angles approaches a uniform distribution, indicating a lack of preferred orientation. This contributes to a dilution of the overall alignment signal when considering the full galaxy population. In contrast, high-mass galaxies tend to preserve a stronger alignment between the jet direction and galaxy orientation.
It is worth noting that our high-mass sample is largely represented by Milky Way–mass galaxies. None of the brightest cluster galaxies (BCGs), for which the misalignment is expected to increase again \citep{dubois+14a,bustamante+19}, are included in our sample.

Finally, Fig.~\ref{fig_dPA_minmax} indicates a transition at stellar masses between 
between 10$^9$M$_\odot$ and 10$^{10}$M$_\odot$, marking a shift from chaotic to rotationally supported galaxies. It is interesting to note that this finding is in line with recent simulations that report a morphological transition occurring at a similar mass scale
\citep[e.g.,][]{stern+21,yu21,yu+23,hafen+22,gurvich+23,hopkins+23w,benavides+25}.

\section{Summary and conclusions}
\label{sec:conclusions}

The orientation of  AGN jet is a crucial probe of the interplay between black hole spin, accretion physics, and host galaxy structure. Whether the BH spin and/or the AGN jet are aligned with the angular momentum of the host galaxy's stellar or gaseous disk remains an open question that holds implications for galaxy formation, black hole co-evolution, and feedback processes.
Using the \nhs simulation, we performed a statistical
  analysis of the projected angles between the BH spin and the host
  galaxy rotation axes.  In order to make a testable prediction for
  photometric and/or spectroscopic surveys of galaxies, we created
  mock catalogs of synthetics DESI-LS and Euclid optical images
  ($r$-band) and synthetic MaNGA-like velocity maps for 100 simulated
  BH-galaxy systems viewed from 50 random line-of-sight directions.
  Then, we computed the optical and kinematic positions angles derived
  from the photometric images and velocity maps, respectively.
  Assuming that the BH spin axis is a good proxy of the observable AGN
  jet direction, we statistically examined the projected angles between the
  the BH spin direction and the host galaxy position angles.
  Our major findings are summarized as follows:

{\bf 1.} 
Both optical and kinematic position angles can be used as
statistically reliable estimators for the projected orientation of the
angular momentum axis of galaxies.  For those systems with
accurate morphological or kinematic measurements, approximately
90\% and 95\% of galaxies exhibit \dPAgalopt<20\degrees\, and  
\dPAgalkin<20\degrees\,, respectively.

{\bf 2.} 
Even after taking account of the possible
  misalignment of the BH spin and the jet directions, we found that
  the projected AGN jets and the position angles of their host
  galaxies exhibit a statistically significant alignment trend.
These findings are in good agreement with recent
observational studies that report similar alignment trends, including
those by \cite{zheng+24}, \cite{fernandez+25} and \cite{jung+25},
which use combinations of radio, optical, and spectroscopic data to
probe jet-host alignment in nearby AGN systems.

{\bf 3.} 
Future
analyses incorporating kinematic position angles from
surveys like MaNGA are very promising to test and
refine the alignment trends identified in this work.

We note that kinematic position angles can be derived not only from
the stellar component but also from the ionized gas, as measured by
integral field unit surveys such as CALIFA or MaNGA. Including gas kinematics
provide additional constraints on jet–galaxy misalignment, as gas may
respond differently to AGN-driven outflows or environmental
effects \citep[see, for instance,][]{barrera-ballesteros+15, garay-solis23,garay-solis+24,garay-solis+25}.

The potential dependence of jet–galaxy misalignment  on the large-scale cosmic environment, such as proximity to cosmic filaments, has been explored in recent studies \citep[e.g.,][]{jung+25}. However, investigating such environmental effects is beyond the scope of the present work. Our current analysis is indeed based on a relatively small sample of 100 BH–galaxy systems, which limits our ability to draw statistically robust conclusions about environmental trends. Nonetheless, this represents a promising avenue for future research. Expanding the sample size and incorporating environmental metrics could provide valuable insights into how cosmic structure influences the orientation and evolution of AGN jets relative to their host galaxies.

To conclude, the tendency for alignment between AGN jets and the spin axes of their host galaxies, suggested by both theoretical predictions and growing observational evidence,
 may have important implication regarding the co-evolution 
of BH and host galaxy, as well as model of galaxy formation in general.
Future high-resolution radio surveys 
such as the ngVLA \citep[the Next Generation Very Large Array,][]{ngVLA} 
or SKAO \citep[Square Kilometre Array Observatory,][]{skao}
combined with integral field spectroscopy, in particular, MaNGA \citep[Mapping Nearby Galaxies at Apache Point Observatory,][]{manga} 
and
MUSE \citep[Multi Unit Spectroscopic Explorer,][]{muse},
will enable statistically robust, spatially resolved measurements of jet orientation and galaxy spin across cosmic time. On the theoretical side, future cosmological simulations that simultaneously resolve both large-scale environments and sub-parsec accretion physics, including magnetohydrodynamic and radiative feedback processes, will be crucial to understanding the origin and persistence of jet–galaxy alignment in different evolutionary contexts.

\begin{acknowledgements}
We warmly thank the referee for their insightful review and valuable comments.
This research is partly supported by the JSPS KAKENHI grant Nos. 23H01212 (Y.S.).
CL acknowledges the support of the French Agence Nationale de la Recherche (ANR), under grant ANR-22-CE31-0007 (project IMAGE).
This work was granted access to the HPC resources of CINES under the allocations c2016047637, A0020407637 and A0070402192 by Genci, KSC-2017-G2-0003, KSC-2020-CRE-0055 and KSC-2020-CRE-0280 by KISTI, and as a “Grand Challenge” project granted by GENCI on the AMD Rome extension of the Joliot Curie supercomputer at TGCC. The large data transfer was supported by KREONET, which is managed and operated by KISTI. S.K.Y. acknowledges support from the Korean National Research Foundation (RS-2025-00514475; RS-2022-NR070872).
This work was carried within the framework of the
Horizon project (\href{http://www.projet-horizon.fr}{http://www.projet-horizon.fr}).
Most of the numerical modeling presented here was done on the Horizon cluster at Institut d'Astrophysique de Paris (IAP).
\end{acknowledgements}

\bibliographystyle{aa}
\bibliography{author}

\end{document}